\global\def\draftcontrol{0}

   \def\versionno{ sbrane -- draft -- 13.08.02   }

\catcode`\@=11

\expandafter\ifx\csname draftcontrol\endcsname\relax\global\def\draftcontrol{0}
\fi

{\count255=\time\divide\count255 by 60
\xdef\hourmin{\number\count255}
\multiply\count255 by-60\advance\count255 by\time
\xdef\hourmin{\hourmin:\ifnum\count255<10 0\fi\the\count255}}
\def\draftdate{\number\month/\number\day/\number\year\ \ \ \hourmin }

\newcommand\makepapertitle{\par
  \begingroup
    \renewcommand\thefootnote{\@fnsymbol\c@footnote}%
    \def\@makefnmark{\rlap{\@textsuperscript{\normalfont\@thefnmark}}}%
    \long\def\@makefntext##1{\parindent 1em\noindent
            \hb@xt@1.8em{%
                \hss\@textsuperscript{\normalfont\@thefnmark}}##1}%
     \newpage
     \global\@topnum\z@   
     \@makepapertitle
     \thispagestyle{empty}\@thanks
  \endgroup
  \setcounter{footnote}{0}%
  \global\let\thanks\relax
  \global\let\makepapertitle\relax
  \global\let\@makepapertitle\relax
  \global\let\@thanks\@empty
  \global\let\@author\@empty
  \global\let\@date\@empty
  \global\let\@title\@empty
  \global\let\title\relax
  \global\let\author\relax
  \global\let\date\relax
  \global\let\and\relax
  \def\version{\let\version\@version\@gobble}
}
\def\@makepapertitle{%
  \newpage
   \ifnum\draftcontrol=1 {}
   \version\versionno
   \vskip 3em%
   \else
   \hfill\hbox to 3cm {\parbox{4cm}{\@pubnum}\hss}%
   \vskip 3em%
   \fi
   \begin{center}%
   \let \footnote \thanks
     {\LARGE {\@title}}%
     \vskip 1.5em%
     {\normalsize
       \lineskip .5em%
       \begin{tabular}[t]{c}%
         \@author
       \end{tabular}\par}%
     \vskip 1.5em%
     {\@bstract}%
     \end{center}%
     \vskip 1.5em 
     \@date%
   \par
}

\gdef\@pubnum{}
\def\pubnum#1{%
  \gdef\@pubnum{#1}}

\gdef\@bstract{}
\def\Abstract#1{%
  \gdef\@bstract{%
   \parbox{\textwidth-0pc}{%
   \centerline{\bf Abstract}\penalty1000
   \noindent
   \renewcommand\baselinestretch{1.0}
   {#1}}}
}

\def\ps@paper{\let\@mkboth\@gobbletwo%
     \ifnum\draftcontrol=1
        \def\@oddfoot{\hbox to \textwidth{\tiny \versionno \hfil\tiny\draftdate}%
        \hskip -\textwidth \hbox to \textwidth{\hfil\rm\thepage\hfil}}%
     \else\def\@oddfoot{\hbox to \textwidth{\hfil\rm\thepage\hfil}}
     \fi
     \let\@evenfoot\@oddfoot
}

\def\body{\clearpage
          \pagestyle{paper}
        }
\newenvironment{acknowledgments}{%
\vskip 3.25ex
\noindent {\bf Acknowledgments}
}

\def\@version#1{\ifnum\draftcontrol=1
\typeout{}\typeout{#1}\typeout{}
\vskip3mm\centerline{\hbox{\fbox{\normalsize{\tt DRAFT -- #1 -- }
                   {\draftdate}}}}\vskip3mm
\fi}
\let\version\@version
\long\def\eqlabel#1{\ifnum\draftcontrol=1
                    \tag@false  
                    \tag*{(\theequation) \hbox to -0.2cm{\hspace{0cm}\small{#1}\hss}}
                    \refstepcounter{equation} 
                    \edef\@currentlabel{\theequation}
                    \ltx@label{#1}          
                    \else
                    \label{#1}
                    \fi
                    }
\let\st@bibitem\@bibitem
\let\st@lbibitem\@lbibitem
\ifnum\draftcontrol=1
  \def\@bibitem#1{%
    \st@bibitem{#1}\a@@label{#1}\ignorespaces}
  \def\@lbibitem[#1]#2{%
    \st@lbibitem[#1]{#2}\a@@label{#2}\ignorespaces}
  \def\a@@label#1{%
    \gdef\a@lab{\smash{\normalfont\small#1}}
    \ifvmode
      \if@inlabel
        \global\setbox\@labels\hbox{%
          \llap{\a@lab\let\a@lab\relax
                \kern\@totalleftmargin\kern\marginparsep}%
          \box\@labels}%
      \fi
    \fi}
\fi

\documentclass[12pt,letterpaper]{article}

\usepackage{amsmath,amssymb,array,calc,rotating,epsfig,psfrag}
\usepackage[nosort]{cite}

\ifnum\draftcontrol=1
\fi

\renewcommand\baselinestretch{1.25}
\renewcommand\section{\@startsection {section}{1}{\z@}%
                                   {-3.5ex \@plus -1ex \@minus -.2ex}%
                                   {2.3ex \@plus.2ex}%
                                   {\normalfont\large\bfseries}}
\renewcommand\subsection{\@startsection{subsection}{2}{\z@}%
                                   {-3.25ex\@plus -1ex \@minus -.2ex}%
                                   {1.5ex \@plus .2ex}%
                                   {\normalfont\normalsize\bfseries}}
\renewcommand\subsubsection{\@startsection{subsubsection}{3}{\z@}%
                                   {-3.25ex\@plus -1ex \@minus -.2ex}%
                                   {1.5ex \@plus .2ex}%
                                   {\normalfont\normalsize\it}}
\renewcommand\paragraph{\@startsection{paragraph}{4}{\z@}%
                                   {-3.25ex\@plus -1ex \@minus -.2ex}%
                                   {1.5ex \@plus .2ex}%
                                   {\normalfont\normalsize\bf}}
\renewcommand\subparagraph{\@startsection{subparagraph}{5}{\z@}%
                                   {-1.25ex\@plus -1ex \@minus -.2ex}%
                                   {0ex \@plus .2ex}%
                                   {\normalfont\normalsize\it}}

\def\ie{{\it i.e.}}

\def\revise#1       {\raisebox{-0em}{\rule{3pt}{1em}}%
                     \marginpar{\raisebox{.5em}{\vrule width3pt\
                     \vrule width0pt height 0pt depth0.5em
                     \hbox to 0cm{\hspace{0cm}{%
                     \parbox[t]{4em}{\raggedright\footnotesize{#1}}}\hss}}}}

\newcommand\nxt[1]  {\\\fnxt#1}

\def\reals        {{\mathbb R}}

\def\del          {\partial}

\def\sqr#1#2{{\vcenter{\vbox{\hrule height.#2pt  
 \hbox{\vrule width.#2pt height#1pt \kern#1pt
 \vrule width.#2pt}\hrule height.#2pt}}}}

\def\a{\alpha}

\def\t{{\triangle}}
\def\e{{\epsilon}}

\def\SO{{\it SO}}
\def\ISO{{\it ISO}}

\catcode`\@=12

\begin{document}


\title{Does the Tachyon Matter?}

\pubnum{%
NSF-ITP-02-65 \\
YITP-SB-02-38 \\
hep-th/0207235}
\date{July 2002}

\author{Alex Buchel$^{1}$\footnote{\tt buchel@kitp.ucsb.edu},  
Peter Langfelder$^{1,2}$\footnote{\tt peter.langfelder@sunysb.edu},
and Johannes Walcher$^{1}$\footnote{\tt walcher@kitp.ucsb.edu} \\[0.4cm]
\it $^{1}$Kavli Institute for Theoretical Physics \\
\it University of California \\
\it Santa Barbara, CA 93106, USA \\[0.2cm]
\it $^{2}$C.\ N.\ Yang Institute for Theoretical Physics \\
\it State University of New York \\
\it Stony Brook, NY 11794-3840, USA 
}

\Abstract{
We study time-dependent solutions of Einstein-Maxwell gravity in four
dimensions coupled to tachyon matter---the Dirac-Born-Infeld Lagrangian
that provides an effective description of a decaying tachyon on an unstable
D-brane in string theory. Asymptotically, the solutions are similar to the
recently studied space-like brane solutions and carry S-brane charge. They
do not break the Lorentzian R-symmetry. We study the tachyon matter as a 
probe in such a background and analyze its backreaction. For early/late 
times, the tachyon field has a constant energy density and vanishing 
pressure as in flat space. On the other hand, at intermediate times,
the energy density of the tachyon diverges and produces a space-like
curvature singularity.
}


\makepapertitle

\body

\version\versionno


It has been argued in \cite{gust} that string theory and related field
theories have solutions corresponding to topological defects with finite
temporal extent---so-called space-like (S-) branes. In string theory or
supergravity, such solutions carry the same RR charge as the familiar 
time-like D-branes. In fact, the basic argument for existence of S-branes 
in string theory comes from the existence of unstable D-brane systems, 
for example the D3-brane in type IIA string theory. The worldvolume of 
such a brane has a tachyon $T$ which is coupled to a RR form by a term 
such as $\int dT\wedge C^{(3)}$. There are configurations of $T$ which
are topologically non-trivial in time and source the corresponding
magnetic flux. The S-brane is the gravitational backreaction of this 
time-dependent process. Explicit supergravity solutions corresponding to 
S-branes have been proposed in \cite{gust,cgg,kmp}. S-branes are likely 
to provide interesting time-dependent backgrounds of string theory. 
Moreover, due to their kinship with D-branes, they can play a role in 
temporal holography, see \cite{gust} and references therein.

In perturbative string theory, S-branes are described by imposing
Dirichlet boundary conditions on the time coordinate of the string
\cite{gust}. The identification proceeds as for time-like branes through
the RR charge carried by open string boundary conditions. The relation 
of these space-like D-branes to tachyon decay was clarified by Sen 
\cite{sen1}, who showed that open string field theory possesses a family 
of solutions corresponding to the roll of the open string tachyon 
from its maximum down to the closed string vacuum. The rolling tachyon
is peculiar for two reasons. Firstly, the energy density remains localized 
in the plane of the original unstable brane system. This is explained by
the open string origin of the tachyon and the absence of open strings in
the bulk. Secondly, when at late times the tachyon approaches the minimum
of its potential, the tachyon is universally (\ie, independently of the
details of the potential) characterized by a constant energy density
and exponentially vanishing pressure. This state of the string field is 
called tachyon matter \cite{sen2}. Moreover, tachyon matter admits an 
effective description by a Dirac-Born-Infeld type Lagrangian \cite{ghy,sen3}. 
The advantage of this field theory description is that it provides a drastic 
simplification in a complicated initial value problem in open string field 
theory, see \cite{zwiebach,sen4} for recent discussions.  Tachyon 
matter has been derived in the context of BSFT in \cite{terashima2}, and has 
also attracted considerable cosmological interest \cite{tachcosmo}.

In this note, we explore the relation of charged space-like branes with
tachyon decay one step further. In the context presented in \cite{gust},
it is natural to ask for the inclusion of tachyon matter \cite{sen1,sen2,
sen3} into the effective dynamics. In particular, one might hope to
see explicitly the excitation (or decay) of the tachyon by incoming (or into
outgoing) radiation. Furthermore, one would like to know whether the
tachyon matter changes the nature of the singularities that plague the 
explicit supergravity solutions found in \cite{gust,cgg,kmp}. Here, we 
study the modification of these backgrounds produced by coupling tachyon 
matter to the supergravity fields. We find that while the asymptotics
are essentially unchanged, the tachyon indeed modifies the singularity 
structure, producing a space-like curvature in the middle of space-time. 
For simplicity, we will consider Einstein-Maxwell theory in four 
dimensions, which yielded the simplest S-brane example in \cite{gust}, 
but we expect that the qualitative picture is similar in other cases.

\paragraph{S-branes}

We begin with a lightning review of the supergravity properties of
S-branes \cite{gust}.

S-branes are expected to exist as time-dependent solutions both in 
ten-dimensional IIA/IIB supergravity, and in eleven-dimensional 
supergravity. Conventionally, the worldvolume of an S$p$-brane is
a $p+1$-dimensional Euclidean space. The types of charge that are
expected to occur are the same as for D-branes, so that in IIA/IIB
supergravity, we have odd/even codimension S-branes with RR charge, 
as well as SNS5-branes and strings. Furthermore, we have S5-brane 
and S2-brane in $D=11$ supergravity. However, the properties of
S-branes should depend drastically on the parity of the codimension. 
More precisely, as explained in \cite{gust}, even codimension
branes should have fluxes supported {\it on} their lightcone, while
odd codimension S-branes source a flux {\it inside} their entire
lightcone. In explicit supergravity solutions, the former seems
much harder to realize, and unsmeared S-brane solutions in IIB
supergravity might not exist\footnote{It is possible to construct 
time-dependent solutions in type IIB supergravity which have an
S-brane background ansatz analogous to \cite{gust}. We will not 
discuss these solutions here.}. In this paper we consider only odd 
codimension S-branes. 

A flat S$p$-brane in $D$ dimensions has a ``classical'' $\ISO(p\!+\!1)
\times\SO(D\!-\!p\!-\!2,1)$ symmetry. Supergravity solutions preserving 
this symmetry are singular. In \cite{gust} it was argued that the
resolution of these singularities should involve a (spontaneous) 
breaking of the transverse R-symmetry down to $\SO(D\!-\!p\!-\!2)$. 
Indeed, this is precisely the breaking that is realized if one thinks 
of the space-like $p$-brane as a decaying unstable time-like 
$p\!+\!1$-brane.

Let us now turn to the simplest example of a space-like brane, the
S$0$-brane in four-dimensional Einstein-Maxwell gravity \cite{gust}. 
The metric is given by
\begin{equation}
ds^2=-\frac{d\tau^2}{ \lambda^2}+\lambda^2 dz^2 +R^2 dH_2^2 \;,
\eqlabel{mansatz}
\end{equation}
where $dH_2^2$ is the normalized metric on the two-dimensional hyperbolic 
space $H_2$. All warp factors depend on $\tau$ only, and the metric has a 
manifest $\SO(2,1)\times\reals$ symmetry. Moreover, the (electric) flux 
is naturally given by the volume form on $H_2$,
\begin{equation}
\star F=2 Q \epsilon_2 \,,
\eqlabel{fst}
\end{equation}
where $Q$ is the charge of the brane. Integration of the equations of
motion yields \cite{gust} 
\begin{equation}
R^2=\frac{Q^2\tau^2}{\tau_0^2}\;,\qquad\qquad
\lambda^2=\frac{\tau_0^2}{Q^2}\frac{\tau^2-\tau_0^2}{\tau^2} \;.
\eqlabel{strominger}
\end{equation}

\begin{figure}
\begin{center}
\psfrag{infty}{$\tau=-\infty$}
\psfrag{zero}{$\tau=0$}
\psfrag{tzero}{$\tau=\tau_0$}
\epsfig{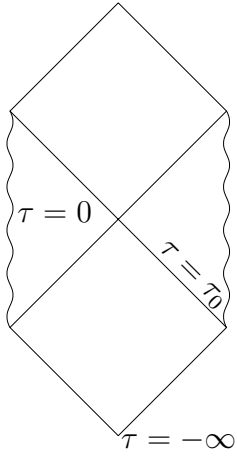}
\caption{Penrose diagram for the metric \eqref{mansatz}. The wavy
lines are time-like curvature singularities.}
\label{pendg}
\end{center}
\end{figure}

One of the characteristics of the solution\footnote{Some of these 
arguments leading to the Penrose diagram in fig.\ \ref{pendg} were
given in \cite{quevedo1,quevedo2}.} \eqref{strominger} is the
vanishing of the $\lambda$ warp factor at $\tau=\tau_0$ at finite
curvature. This is reminiscent of a black hole horizon. Indeed, the 
simplest way to arrive at the solution \eqref{strominger} is to
analytically continue the Reissner-Nordstr\"om black hole at imaginary
charge and zero mass (the mass corresponds to a second integration
constant in \eqref{strominger}, which is set to zero by requiring
$\tau\to -\tau$ symmetry). This also gives an easy way to obtain the 
maximally extended spacetime associated with the metric \eqref{mansatz}.

We have depicted (the projection onto the $\tau$-$z$-plane of) this maximally 
extended spacetime in fig.\ \ref{pendg}. As for a black hole, there are two 
asymptotic regions, separated by a region of strong curvature. Here, the 
two regions are causally connected. Another interesting region is close
to the coordinate singularity at $\lambda=0$, which we will refer to as
the Milne region. In addition, there is a genuine curvature singularity at 
$\tau=0$. This is a naked {\it time-like} (rather than {\it space-like}!)
singularity associated with the vanishing of the scale-factor of the 
transverse spatial directions of the S-brane. It is worthwhile to stress 
that only one of the asymptotic regions corresponds to a piece of the 
metric \eqref{mansatz}, while the other is obtained by continuing past
the Milne singularity. One should note that overall the diagram looks
little like what one would have expected for a space-like brane%
\footnote{We thank Gary Horowitz for emphasizing this.}.

We will now introduce the coupling of the tachyon matter and then study 
its effect on the background \eqref{mansatz}, focusing on the three 
interesting regions in turn.

\paragraph{Coupling the Tachyon Matter}
 
In general, one should be cautious about supergravity (string theory) 
backgrounds containing naked singularities. Certain singularities should 
not be resolvable, and backgrounds containing them should simply be 
disallowed \cite{hm}. However, if the singularity occurs in a background 
for which there is a solid physical argument that it must be consistent, 
we expect that such singularity be resolved by incorporating the right 
physics. Adopting the viewpoint that an S-brane is a time-dependent
description of an unstable D-brane, such branes are well-motivated
physical objects, and we expect that it is possible to give a meaning
to the singularities. 

The definition of S-branes also suggests a natural way to try to make 
progress on understanding the above background, namely by including the 
decaying unstable D-brane as a source. Explicitly, we assume that the 
missing mode is the open string tachyon of the corresponding D-brane. 
The general form of the coupling of this tachyon to supergravity fields 
is quite well-known \cite{sen,billo,garousi,bergshoeff,terashima1}. There 
is a DBI term containing the tachyon kinetic term and potential, and a WZ 
term describing the coupling of the tachyon to the RR fields. The precise 
expressions for the tachyon potential and tachyon dependence of the RR 
coupling are not known in all cases, but most of our results only depend 
on their generic features.

Specifically, the S$0$-brane that we have reviewed above should be the 
decay of an ``unstable D$1$-brane'' in 4-d Einstein-Maxwell gravity.
This can be viewed as a toy model of type IIA string compactification 
with suppressed scalar moduli, and it is this model that we will study in 
the following. The action is given by
\begin{multline}
S=S_{\it EM}+S_{\it branes} \\
={\textstyle\frac1{16\pi G_4}}\int d^4x\,\sqrt{-g}\,
\bigl(R-{\textstyle\frac 14}F^2\bigr)
+\mu_1\int d^4 x\, \rho(x_\perp)\bigl[ -V(T)\,\sqrt{-A} + 
f(T) dT\wedge C^{(1)}\bigr] \,,
\label{staction}
\end{multline}
where $A=\det \bigl(g_{\mu\nu}+\del_{\mu}T\del_{\nu}T\bigr)$. In 
\eqref{staction}, $G_4$ is the four-dimensional Newton constant, $\mu_1$ 
is the tension of the unstable brane, $V(T)$ is the tachyon potential, and
$f(T)$ is the tachyon coupling to the ``RR'' one-form $C^{(1)}$, with flux 
$F=dC^{(1)}$. Moreover, $\rho(x_\perp)$ is the density of D$1$-branes
in the transverse space, further discussed below. In the following, we 
will use Roman indices or the symbol $\perp$ to denote directions 
transverse to the D1-brane, and Greek indices or the symbol $\parallel$ 
for the directions parallel to it, so that for instance,
$g=\det g_{\parallel}\,\det g_{\perp}\equiv\det g_{\mu\nu}\,\det g_{ij}\,$.

The tachyon potential and the RR coupling is not known exactly, but we 
will be mostly concerned with universal features of the system that are 
independent of the exact form of $V$ and $f$. Specifically, we shall 
assume that $V(T)$ is smooth at the unstable maximum, which we take to
be at $T=0$. We also assume symmetries, $V(T)=V(-T),\ f(T)=-f(-T)$, 
and the universal asymptotics \cite{sen2,sen3}
\begin{equation}
V(|T|\to \infty)\to e^{-|T|/\sqrt{2}}\;, \quad\qquad
f(|T|\to \infty)\to {\rm sign}(T)\ e^{-|T|/\sqrt{2}} \;.
\eqlabel{tpot}
\end{equation}  

It is straightforward to derive equations of motion for \eqref{staction}.
We find the following Einstein equations
\begin{equation}
\begin{split}
R_{\mu\nu}&={\frac 14}\Lambda V(T)\rho(x_i)
\sqrt{\frac{A}{g}}
\;\bigl[(A^{-1})^{\alpha\beta}g_{\a\beta}g_{\mu\nu}
-2 (A^{-1})^{\a\beta} g_{\a\mu}g _{\beta\nu}\bigr]+
T^{(2)}_{\mu\nu}\cr
R_{ij}&= {\frac 14} \Lambda V(T)\rho(x_i)
\sqrt{\frac{A}{g}}
\;\bigl[(A^{-1})^{\a\beta} g_{\a\beta}\bigr] \, g_{ij} +
T^{(2)}_{ij} \,,
\end{split}
\eqlabel{sein}
\end{equation}
where $\Lambda\equiv 16\pi G_4 \mu_1\,$, and 
\begin{equation}
T_{AB}={\textstyle\frac 12}\bigl[F_{AC} F_{B}\ ^C-{\textstyle\frac 14} 
F^2 g_{AB}\bigr]
\eqlabel{fluxT}
\end{equation}
is the stress tensor of the 2-form flux $F_{AB}$. We also have a 
Maxwell equation
\begin{equation}
\del_A\bigl(\sqrt{-g} F^{AB}\bigr)+\Lambda f \rho(x_i)
\left[\del_{\alpha}T \delta_{\beta}^B-\del_{\beta}T
\delta_{\alpha}^B\right]=0 \,,
\eqlabel{maxwell}
\end{equation}
and the tachyon equation
\begin{equation}
\rho(x_i)\Bigl( \sqrt{-A}\ \frac{d V}{d T}+fF_{\alpha\beta}\Bigr)
-\del_A\bigl[ \rho(x_i)V
\sqrt{-A} \left(A^{-1}\right)^{AB}\del_B T 
\bigr]=0 \,.
\eqlabel{taceq}
\end{equation}

We are looking for solutions of \eqref{sein}, \eqref{maxwell} and
\eqref{taceq} that in the asymptotic regions look like the solutions 
without tachyon, \ie, eq.\ \eqref{mansatz}. As discussed in \cite{gust}, it 
is likely that in order to resolve the singularity, one must violate the
$\SO(2,1)$ R-symmetry of this solution in the intermediate regions.
Indeed, the most natural way to achieve this in our framework is to
localize the decaying D$1$-branes at a point in the transverse space, \ie, 
to put $\rho(x_\perp)=\delta(x_\perp)$. However, doing so makes the 
equations intractable. We shall instead {\it preserve} the R-symmetry 
by {\it smearing} the branes and tachyon matter {\it uniformly} along 
the transverse space. Explicitly, we take
\begin{equation}
\rho(x_i) = \rho \sqrt{g_\perp} \;,
\eqlabel{smearing}
\end{equation}
where $\rho$ is a constant number density of D$1$-branes. With the ansatz 
\begin{equation}
ds^2=-c_1^2 d\tau^2+c_2^2 dz^2+c_3^2 dH_2^2
\eqlabel{cmetric}
\end{equation}
for the metric, the two-form flux
\begin{equation}
F_{\tau z}=A\ c_1 c_2 \,,
\eqlabel{cflux}
\end{equation}
and with only time-dependent warp-factors $c_i=c_i(\tau)$, $A=A(\tau)$, and 
tachyon $T=T(\tau)$, we obtain the following differential equations 
\begin{equation}
\begin{split}
R_{\tau\tau}
=\frac{1}{c_1 c_2 c_3}\biggl(c_3\Bigl(\frac{c_2'}{c_1}
\Bigr)'+2 c_2 \Bigl(\frac{c_3'}{c_1}\Bigr)'\biggr) g_{\tau\tau}
&=\biggl(\frac{\Lambda V\rho}{4 c_3^2}
\bigl\{\t^{1/2}-\t^{-1/2}\bigr\} -\frac{A^2}{4} \biggr) g_{\tau\tau}\cr
R_{zz}
=\frac{1}{c_1 c_2 c_3^2}\Bigl(\frac{c_3^2 c_2'}{c_1}\Bigr)' g_{zz}
&=-\biggl(\frac{\Lambda V\rho}{4 c_3^2}
\bigl\{\t^{1/2}-\t^{-1/2}\bigr\} +\frac{A^2}{4} \biggr) g_{zz}\cr
R_{ij}
=\frac{1}{c_1 c_2 c_3^2}\biggl(\Bigl(\frac{c_2 c_3 c_3'}{c_1}\Bigr)'
-c_1 c_2\biggr) g_{ij} 
&= \biggl(\frac{\Lambda V\rho}{4 c_3^2}
\bigl\{\t^{1/2}+\t^{-1/2}\bigr\} +\frac{A^2}{4} \biggr) g_{ij}
\end{split}
\eqlabel{diff} 
\end{equation}
\begin{equation}
\left(\frac{d V}{d T}\t^{1/2}+\frac{d F }{d T} A \right)c_1 c_2 +\left[
\frac{T' V\t^{-1/2} c_2}{c_1}
\right]' =0 \,,
\eqlabel{tahcequa}
\end{equation}
where we introduced 
\begin{equation}
\t\equiv 1-\frac{(T')^2}{c_1^2}\;,\qquad\qquad
\frac {d F(T)}{d T}\equiv f(T) \,.
\eqlabel{trian}
\end{equation}
The Maxwell equation \eqref{maxwell} can be explicitly integrated with 
the result
\begin{equation}
A=\frac{2 Q+\Lambda \rho F}{c_3^2}\;,
\eqlabel{inta}
\end{equation} 
where the integration constant $Q$ has been chosen in such a way as
to reproduce the S0-brane solution of \eqref{mansatz}-\eqref{strominger}
as $\Lambda=0$.

We now proceed to solving eqs.\ \eqref{diff} and \eqref{tahcequa}, using
\eqref{inta}. We will first discuss asymptotics at late times, then
close to a Milne coordinate singularity and finally near a curvature
singularity. Both types of singularities appear in \eqref{strominger},
and we wish to see whether including the tachyon changes their
structure. In what follows, we will usually only need the generic
behavior \eqref{tpot}. For some purposes, such as numerical integration, 
for example, it is useful to have an explicit expression for $V$ and $f$. 
Whenever appropriate, we have used
\begin{equation}
V(T)=\frac{1}{\cosh T/\sqrt{2}} \;,\qquad\quad
F(T)=-\frac{\sqrt{2}}{\cosh T/\sqrt{2}} \;.
\eqlabel{potoupl}
\end{equation}

\paragraph{Asymptotic Region}

Tachyon condensation in a flat background produces pressureless gas with
constant energy density \cite{sen1,sen2,sen3}. This non-zero energy density
is the basic reason for asking for the backreaction of the tachyon on 
S-brane backgrounds. One might expect in particular that the asymptotic 
flatness of the space is modified. In our model, the tachyon still produces a
pressureless gas at late times, but the energy density is ``time-diluted''
because of the smearing of the branes. We find that the backreaction
produces a logarithmic correction to flat space.
 
\subparagraph{Tachyon matter in asymptotic S0-brane background}\

\noindent
We begin by treating the tachyon matter as a probe in the background
\eqref{strominger} in the region $\tau\to\infty$. We will see that 
$T\to \infty$, so we can use the asymptotics \eqref{tpot} for the tachyon 
potential and the coupling to the RR potential. The tachyon equation 
becomes
\begin{equation}
0\approx 2 Q\t(\tau)\sqrt{1-\t(\tau)}+\sqrt{2}\tau_0\frac{d\t(\tau)}{d\tau}\,,
\eqlabel{inftyass}
\end{equation}
with the solution 
\begin{equation}
\t=\cosh^{-2}\left(\frac{Q\tau}{\sqrt{2}\tau_0}+\beta\right) \,,
\eqlabel{tsolve}
\end{equation}
where $\beta$ is an integration constant. From \eqref{tsolve} and 
\eqref{trian} it follows that asymptotically the tachyon probe behaves as 
\begin{equation}
T=\frac{Q\tau}{\tau_0}+\frac{1}{2^{3/2}}e^{-2\left(\frac{Q\tau}{\sqrt{2}
\tau_0}+\beta\right)}+o(e^{-\frac{\sqrt{2} Q\tau}{\tau_0}}) \,.
\eqlabel{tachprobel}
\end{equation}
Notice that at infinity, $V \t^{-1/2}\to {\rm const}$, rather similar
to the flat space result \cite{sen3}. The only difference is that,
because of the smearing, the energy density is here diluted,
$\rho_{\it tachyon}\sim \tau^{-2}$. The pressure still decays exponentially
$P_{\it tachyon}\sim e^{-\sqrt{2}Q \tau/\tau_0}/\tau^2$. Since the 
flux contribution to the stress tensor behaves asymptotically like 
$T^{(2)}\sim \tau^{-4}$, the tachyon seems to dominate at late times.
We now show that this basic conclusion is unchanged if we include the
backreaction.
   
\subparagraph{Backreaction}\

\noindent
In eqs.\ \eqref{diff} and \eqref{tahcequa}, let us neglect all exponentially
suppressed terms, \ie, we neglect $\t^{1/2}$ over $\t^{-1/2}$, and $F$ over
$Q$ for the gauge potential \eqref{inta}. Choosing the gauge where
$c_3=Q\tau/\tau_0$, a certain combination of Einstein equations reads
\begin{equation}
0=Q^2 \tau^2 (2\tau c_1'-c_1)+\tau_0^2 c_1^3 (\tau^2+\tau_0^2) \,,
\eqlabel{assc1}
\end{equation}
which can be readily integrated to yield
\begin{equation}
c_1^2=\frac{Q^2}{\tau_0^2}\ \frac{\tau^2}{\tau^2-\tau_0^2+\beta\tau} \,,
\eqlabel{c1result}
\end{equation}
where $\beta$ is an integration constant. Notice that this warp
factor is the same as in the original S$0$-brane solution 
\eqref{strominger}, where the integration constant $\beta$ was set
to zero. Using this, we can write the Einstein equations in terms of 
$\e(\tau)\equiv V \t^{-1/2}\,$,
\begin{equation}
\begin{split}
\e'&=-\frac{\e (\Lambda\rho \tau^2 \e +4\tau_0^2-2\beta\tau)}
{4\tau(\tau^2-\tau_0^2+\beta\tau)} \cr
0&=\left(\e c_2\right)' \;.
\end{split}
\eqlabel{2ein}
\end{equation}
The latter equations can be solved analytically with the result
\begin{equation}
\begin{split}
\e^{-1}&=\frac{1}{4\tau}\biggl\{\Lambda\rho\sqrt{\tau^2-\tau_0^2
+\beta\tau}\ln\left[
\beta_1 (\beta+2\tau+2\sqrt{\tau^2-\tau_0^2+\beta\tau})\right]\cr
&\quad+\frac{2\Lambda\rho(\tau_0^2\beta-2\tau\tau_0^2-\tau\beta^2)}{4\tau_0^2+
\beta^2}\biggr\}\cr
c_2&=\beta_2/\e \,,
\end{split}
\eqlabel{solutionc2e}
\end{equation}
where $\beta_1,\beta_2$  are additional integration constants. We see
that at late times, we have $\e=V \t^{-1/2}\sim 1/\ln\tau$. Therefore,
$\rho_{\it tachyon}\sim 1/(\tau^2\ln\tau)$, and still dominates the 
asymptotic flux stress tensor. And indeed, the warp factor
$c_2\sim \e^{-1}\sim \ln\tau$ as $\tau\to +\infty$. The curvature still
vanishes asymptotically.

\paragraph{Milne Region}

We now consider the effect of tachyon matter on the Milne coordinate 
singularity of the S0-brane solution, which is near $\tau=\tau_0$ in eq.\ 
\eqref{strominger}. We will find that in general the Milne singularity is
in fact {\it replaced} with a genuine space-like curvature singularity.

We begin by exhibiting an analytical solution with static tachyon
$dT/d\tau\equiv 0$. It is convenient to choose the gauge 
$c_1=1/c_2\equiv 1/\lambda$. The coordinate singularity then occurs 
at $\tau=\tau_*$, where $\lambda(\tau_*)=0$, provided that $c_3\equiv 
R(\tau_*)\ne 0$. Note that for a stationary tachyon, which is a solution 
of \eqref{tahcequa} only if tachyon sits at the top of the potential 
$T(\tau)\equiv 0$, we have $\t\equiv 1$. Denoting 
$F\equiv F|_{\tau=\tau_*}=F_0$, $V\equiv V|_{\tau=\tau_*}=V_0$, 
we find an exact analytical solution of the Einstein equations \eqref{diff}
\begin{equation}
\begin{split}
R^2&=\frac{Q^2\tau^2}{\tau_0^2}\cr
\lambda^2&=\frac{\tau_0^2}{Q^2}\left[
\frac{\Lambda V_0 \rho}{2}+\left(1-\frac{\tau_0^2}
{\tau^2}\left(1+\frac{\Lambda \rho F_0}{2 Q}\right)^2\right)+\frac{\beta}
{\tau}\right] \,,
\end{split}
\eqlabel{solmilne}
\end{equation} 
where  $\beta$ is an integration constant. We have chosen the other two 
integration constants such as to reproduce solution \eqref{strominger} as 
$\Lambda=0$. From \eqref{solmilne}, it is easy to see that $\lambda^2$ 
always has a simple zero for a certain $\tau=\tau_* >0$, and everything
is perfectly smooth. Note that although \eqref{solmilne} was obtained for 
a stationary tachyon, it is clearly valid whenever the tachyon evolution 
across the Milne singularity satisfies $|T'(\tau_*)|<\infty$. Indeed, in 
this case $\t=1$ at $\lambda=1$, and thus \eqref{solmilne} is an 
approximate solution. The tachyon acts somewhat like a cosmological 
constant along the decaying D$1$-brane.

We now argue that finiteness of $T'$ is in fact not generic, and that
we rather have $T'|_{\tau\to \tau_*}\to \infty$. To see this%
\footnote{We are here assuming that the tachyon itself has a finite 
expectation value at the Milne coordinate singularity. This is confirmed 
by numerical computation and asymptotic analysis.}, take
$\lambda^2\approx\lambda_0^2 (\tau-\tau_*)$ and $R\approx Q\tau_*/\tau_0$.
Eq.\ \eqref{tahcequa} then yields
\begin{equation}
0 = \t^2-\t+ x \t' \,,
\eqlabel{epeq}
\end{equation}
where we introduced $x\equiv \tau-\tau_*$, and neglected subdominant 
terms as $x\to 0$. The derivative in eq.\ \eqref{epeq} is with respect to 
$x$. The general solution to this equation is 
\begin{equation}
\t=\frac{x}{x+\beta_1}=\frac{\tau-\tau_*}{\tau-\tau_*+\beta_1} \,,
\eqlabel{teqa}
\end{equation}
where $\beta_1$ is a new integration constant. Note that, unless 
$\beta_1=0$, \eqref{teqa} implies that $T'\sim 1/\sqrt{\tau-\tau_*}$
and $T|_{\tau\to \tau_*}\to {\rm const}$. Thus, in this case the energy 
density of the tachyon matter diverges as $\rho_{\it tachyon}\sim V\t^{-1/2}
\sim V_0/\sqrt{\tau-\tau_*}$, producing a space-like curvature singularity 
in the metric. The warp factors can be solved asymptotically to yield
\begin{equation}
\begin{split}
c_2&\approx\lambda_0\sqrt{\tau-\tau_*}\cr
c_1&\approx c_2^{-1}\left(1+\frac{\Lambda V_0\rho \tau_0^2 \sqrt{\tau-\tau_*}}
{Q^2 \tau_*^2\lambda_0^2}+O({\tau-\tau_*})\right)\cr
c_3&\approx \frac{Q \tau_*}{\tau_0}+
\frac{\Lambda V_0\rho \tau_0 \sqrt{\tau-\tau_*}}{Q \tau_*\lambda_0^2}
+O({\tau-\tau_*}) \,.
\end{split}
\eqlabel{metricsin}
\end{equation}

Notice that to obtain \eqref{metricsin} as a solution of Einstein equations 
\eqref{diff}, we had to relax the gauge condition $c_1 c_2\equiv 1$. 
We also note that the special case $\beta_1=0$ in \eqref{teqa} corresponds 
to a smooth evolution of the tachyon across the Milne coordinate 
singularity. The local geometry there is analogous to \eqref{solmilne}.

Another special class of interesting solutions can be obtained (numerically) 
by tuning the tachyon so that it ends up precisely at the top of its potential
at $\tau=\tau_*$. However, because $T'$ (and the curvature) still diverges 
there, it is not clear whether these solutions can be continued by 
``time-reversal'' to the other side. If this were true, the tachyonic kink 
corresponding to the S-brane would have infinite slope at the origin.

We would like to emphasize that our result about destabilization of 
the Milne coordinate singularity to a genuine space-like singularity is 
rather generic and does not depend on the specific form of the tachyon 
potential or the RR coupling. Physically, this is a very satisfying result 
as S-branes should really be space-like defects, and thus
their gravity description should have a space-like singularity. We should
also point out that the divergence of the $\rho_{\it tachyon}$ is not
due to the smearing of the branes. This could matter only if $c_3\to 0$.

\paragraph{Tachyon near time-like singularity}

The original S-brane solution \eqref{strominger} had a time-like singularity 
at $\tau=0$ in the Rindler wedges of the Milne coordinate singularity
(the wavy lines in fig.\ \ref{pendg}). In the previous section we argued that
the coordinate singularity is generically destabilized to a cosmological 
singularity by the tachyon matter. It thus seems really unjustified to
continue past it. Nevertheless, in order to complete the analysis of eqs.\
\eqref{diff}, \eqref{tahcequa}, we have investigated how the tachyon matter 
would modify this time-like singularity. This might also be relevant if 
for very ``finely-tuned''tachyon matter the Milne coordinate singularity 
survives.

The basic result is that the tachyon matter does not help to remove 
or significantly modify the time-like singularity associated with 
vanishing transverse scale factor. More precisely, using only 
the universal properties of the tachyon potential and the RR 
coupling, we have analyzed the asymptotics ``inside'' of the Milne
region, where $\tau$ is really a space-like coordinate. Taking the 
gauge $c_2=1/c_1\equiv \lambda$ and denoting the transverse scale
factor by $R\equiv c_3$, and using the properties of $V$ and $f$ 
discussed above, a power series analysis near $T(\tau=0)=0$, shows the 
following facts.
\nxt
There is no smooth solution with the parity properties $R(-\tau)=
R(\tau)$, $R(\tau=0)\ne 0$, $\lambda(-\tau)=\lambda(\tau)$, 
$\lambda(\tau=0)\ne 0$, and $T(\tau)=-T(-\tau)$. In other words, there 
is no smooth cosmological bounce solution.
\nxt
There is no smooth solution with the parity properties $R(-\tau)=
-R(\tau)$, $\lambda(-\tau)=\lambda(\tau)$, $\lambda(\tau=0)\ne 0$, and 
$T(\tau)=-T(-\tau)$. 
\nxt
There is no smooth solution with the parity properties $R(-\tau)=
-R(\tau)$, $\lambda(-\tau)=-\lambda(\tau)$, and $T(\tau)=-T(-\tau)$.
In other words, the time-like singularity can not coincide with the 
Milne coordinate singularity.

We have confirmed the above results by numerical integration. We have
also studied numerically a few other solutions of the equations
\eqref{diff} and \eqref{tahcequa} that do not fit with any asymptotics 
expected from S-brane solutions. For example, one can look for solutions 
that do not respect any $\tau\to-\tau$ symmetry, but where the
tachyon reaches the top of its potential before the scale factor $R$
vanishes. We found that there are no smooth solutions of this type.
Instead, the curvature always diverges, due to diverging stress tensor
of either the flux or the tachyon.

\paragraph{Discussion and Final Comments}

In this paper, we have analyzed a toy model for understanding the
effect of the decaying tachyon on the supergravity background 
corresponding to space-like branes in string/M-theory. The model was 
obtained by coupling tachyon matter to Einstein-Maxwell gravity in 
four dimensions. Without breaking the R-symmetry, we have studied
the backreaction of tachyon matter on the simplest S$0$-brane model. 
We have found that the tachyon matter modifies slightly the asymptotics
at late/early times, and has a rather drastic effect on the behavior
at intermediate times, where the tachyon matter produces a space-like
singularity which was absent in the original background. The tachyon
does not help to resolve the time-like singularity.

Our results are rather generic. They do not depend, for the most part, 
on the precise details of the tachyon couplings to the gravity fields. 
Furthermore, it is not unreasonable to expect the qualitative picture
to be similar for other space-time and brane dimensionalities.
We have also argued that the preservation of the R-symmetry should
not be crucial for our results, although more detailed investigations
would be needed to confirm this.

To complement the picture, it would be interesting to repeat the
computations for other dimensions and more general brane systems. In 
particular, in higher dimensions, it will be interesting to see how much 
the coupling to the dilaton changes the details of the picture. 
Furthermore, one can imagine similar computations for tachyons in 
D$\overline{\rm D}$-systems. In this context, S-branes correspond to 
space-like tachyonic vortices in higher-dimensional branes. Since the 
general form of the DBI action and RR coupling are also known for such 
systems \cite{sen,billo,garousi,bergshoeff,terashima1}, it should be possible 
to study the backreaction of tachyon matter also on these S-branes.

\begin{acknowledgments}
It is a pleasure to thank Justin David, Oliver DeWolfe, Gary Horowitz,
Kirill Krasnov, Joe Polchinski, and Radu Roiban for interesting and 
stimulating discussions. We would like to thank Justin David for a
critical reading of the manuscript. The research of A.B.\ was supported 
in part by the NSF under Grant Nos.\ PHY00-98395 and PHY99-07949. The 
research of P.L.\ and J.W.\ was supported in part by the NSF under Grant 
No.\ PHY99-07949. P.L.\ would like to thank the KITP at Santa Barbara 
for hospitality while this work was being done.
\end{acknowledgments}


\begin{thebibliography}{99}

\bibitem{gust}
M.~Gutperle and A.~Strominger,
``Spacelike branes,''
JHEP {\bf 0204}, 018 (2002)
[arXiv:hep-th/0202210].

\bibitem{sen1}
A.~Sen
``Rolling tachyon,''
JHEP {\bf 0204}, 048 (2002)
[arXiv:hep-th/0203211].

\bibitem{sen2}
A.~Sen
``Tachyon matter,''
arXiv:hep-th/0203265.

\bibitem{sen3}
A.~Sen
``Field theory of tachyon matter,''
arXiv:hep-th/0204143.

\bibitem{terashima2}
S.~Sugimoto and S.~Terashima,
``Tachyon matter in boundary string field theory,''
JHEP {\bf 0207}, 025 (2002)
[arXiv:hep-th/0205085].

\bibitem{cgg}
C.~M.~Chen, D.~V.~Gal'tsov and M.~Gutperle,
``S-brane solutions in supergravity theories,''
arXiv:hep-th/0204071.

\bibitem{kmp}
M.~Kruczenski, R.~C.~Myers and A.~W.~Peet,
``Supergravity S-branes,''
JHEP {\bf 0205}, 039 (2002)
[arXiv:hep-th/0204144].

\bibitem{tachcosmo}
G.~W.~Gibbons,
``Cosmological evolution of the rolling tachyon,''
Phys.\ Lett.\ B {\bf 537}, 1 (2002)
[arXiv:hep-th/0204008]; \ \
%
M.~Fairbairn and M.~H.~Tytgat,
``Inflation from a tachyon fluid?,''
arXiv:hep-th/0204070;
%
S.~Mukohyama,
``Brane cosmology driven by the rolling tachyon,''
Phys.\ Rev.\ D {\bf 66}, 024009 (2002)
[arXiv:hep-th/0204084];
%
A.Feinstein,
``Power-law inflation from the rolling tachyon''
arXiv:hep-th/0204140;
%
T.~Padmanabhan,
``Accelerated expansion of the universe driven by tachyonic matter,''
Phys.\ Rev.\ D {\bf 66}, 021301 (2002)
[arXiv:hep-th/0204150];
%
G.~Shiu and I.~Wasserman,
``Cosmological constraints on tachyon matter,''
Phys.\ Lett.\ B {\bf 541}, 6 (2002)
[arXiv:hep-th/0205003];
%
G.~Shiu, S.~H.~Tye and I.~Wasserman,
``Rolling tachyon in brane world cosmology from superstring field theory,''
arXiv:hep-th/0207119.

\bibitem{quevedo1}
C.~Grojean, F.~Quevedo, G.~Tasinato and I.~Zavala C.,
``Branes on charged dilatonic backgrounds: Self-tuning, Lorentz  violations and cosmology,''
JHEP {\bf 0108}, 005 (2001)
[arXiv:hep-th/0106120].

\bibitem{quevedo2}
C.~P.~Burgess, F.~Quevedo, S.~J.~Rey, G.~Tasinato and C.~.~Zavala,
``Cosmological spacetimes from negative tension brane backgrounds,''
arXiv:hep-th/0207104.

\bibitem{sen}
A.~Sen,
``Non-BPS states and branes in string theory,''
arXiv:hep-th/9904207.

\bibitem{billo}
M.~Billo, B.~Craps and F.~Roose,
``Ramond-Ramond couplings of non-BPS D-branes,''
JHEP {\bf 9906}, 033 (1999)
[arXiv:hep-th/9905157].

\bibitem{garousi}
M.~R.~Garousi,
``Tachyon couplings on non-BPS D-branes and Dirac-Born-Infeld action,''
Nucl.\ Phys.\ B {\bf 584}, 284 (2000)
[arXiv:hep-th/0003122].

\bibitem{bergshoeff}
E.~A.~Bergshoeff, M.~de Roo, T.~C.~de Wit, E.~Eyras and S.~Panda,
``T-duality and actions for non-BPS D-branes,''
JHEP {\bf 0005}, 009 (2000)
[arXiv:hep-th/0003221].

\bibitem{terashima1}
T.~Takayanagi, S.~Terashima and T.~Uesugi,
``Brane-antibrane action from boundary string field theory,''
JHEP {\bf 0103}, 019 (2001)
[arXiv:hep-th/0012210].

\bibitem{ghy}
G.~W.~Gibbons, K.~Hori and P.~Yi,
``String fluid from unstable D-branes,''
Nucl.\ Phys.\ B {\bf 596}, 136 (2001)
[arXiv:hep-th/0009061].

\bibitem{zwiebach}
N.~Moeller and B.~Zwiebach,
``Dynamics with infinitely many time derivatives and rolling tachyons,''
arXiv:hep-th/0207107.

\bibitem{sen4}
A.~Sen,
``Time evolution in open string theory,''
arXiv:hep-th/0207105.

\bibitem{hm}
G.~T.~Horowitz and R.~C.~Myers,
``The value of singularities,''
Gen.\ Rel.\ Grav.\  {\bf 27}, 915 (1995)
[arXiv:gr-qc/9503062].


\end{thebibliography}
\end{document}